\documentstyle[11pt,moriond,epsfig]{article}

\def\ra{\rightarrow}

\def\be{\begin{equation}}
\def\ee{\end{equation}}
\def\bea{\begin{eqnarray}}
\def\eea{\end{eqnarray}}

\begin{document}
\vspace*{4cm}
\title{STANDARD MODEL HIGGS SEARCH AT LEP IN CHANNELS OTHER THAN FOUR JETS}

\author{ E. PIOTTO }

\address{EP Division, CERN, CH-1211 Geneve 23, Switzerland.}

\maketitle\abstracts{The LEP centre of mass energy has been increased since 1996 in the aim
of producing the Higgs boson. The SM Higgs boson search has been pursued in the four LEP collaborations
exploiting final states with higher branching ratios. In the following we discuss 
the search in final states with two jets and missing 
energy or charged leptons.}

\section{Introduction}

In the year 2000 the four LEP experiments collected data at centre of mass energies between 200
and 209 GeV, integrating approximately 870 pb$^{-1}$ of luminosity. The results presented here 
are based on the analysis published soon after the end of data 
taking~\cite{aleph}$^-$~\cite{opal}.

At LEP the Higgs is expected to be produced mainly via Higgstrahlung process $e^+e^-\ra HZ^0$
and via the WW-fusion: $e^+e^-\ra WW\nu _e\nu _e\ra H\nu _e\nu _e $ which has nearly the same
contribution at kinematic  limit $m(H)=\sqrt{s}-m(Z^0)$. The contribution
from the ZZ-fusion process is an order of magnitude smaller than the WW-fusion process.

For the masses accessible at LEP the Higgs decays mainly into b quarks.
For an Higgs mass of 115 GeV/c$^2$ the BR($H\ra b\overline{b}$) is 78\% and 
the BR($H\ra \tau^+\tau^-$) is 7.5\%.
Thus the Higgs search is divided into channels with final states characterised
by the decays of the $Z^0$: 4-jets ($HZ^0\ra b\overline{b}q\overline{q}$) for 56\% of the 
cases; 2-jets and missing energy ($HZ^0\ra b\overline{b}\nu\overline{\nu}$) for 16\%;
leptonic ($HZ^0\ra b\overline{b}e^-e^+$ and $HZ^0\ra b\overline{b}\mu^-\mu^+$) for 5\% 
and taus ($HZ^0\ra b\overline{b}\tau^-\tau^+$ and $\tau^-\tau^+b\overline{b}$)
 for 8\% of the cases.

\begin{figure}[hbtp]
\begin{center}
\mbox{\epsfig{file=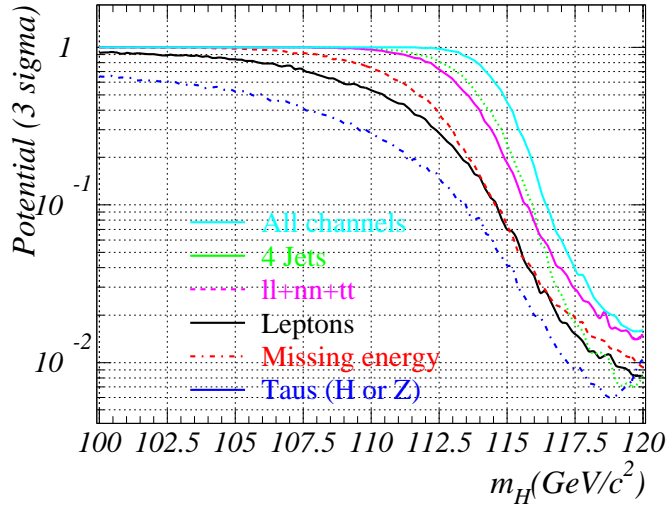,width=0.6\textwidth}}
\caption{3 sigma search potential for different final states.
}
\label{potential}
\end{center}
\end{figure}

Due to the different background composition and mass  resolution,
the missing energy final state together with the leptonic and taus final states have the same search 
potential as the four jets final state (see figure \ref{potential}) despite the branching ratio
being nearly one half. 

\section{The search strategy}
The cross section for Higgs of 115 GeV/c$^2$ is 0.058 picobarn at 207 GeV centre of mass energy,
while the background processes have
cross sections of the order of several tens of picobarn ($e^-e^+\ra q\overline{q}\gamma$,
$e^-e^+\ra WW$ ) or even several hundreds of picobarn ($e^-e^+\ra e^-e^+$X).
The strategy followed by the LEP collaborations to select the signal is based on two step
analysis: firstly a series of sequential cuts is applied to reject the bulk of background,
secondly a likelihood  or a Neural Network is used to profit from the different 
kinematic   distributions of signal and background.  

\subsection{The sequential cuts}
In the process $e^+e^-\ra e^+e^-\gamma\gamma\ra e^+e^-X$, electron and positron of the 
final state escape detection since they go down the beam pipe. As a consequence the total
visible energy in the plane perpendicular to the beam line is low and the spectrum of visible
invariant mass is concentrated at low values.
In all Higgs search  it is possible to reduce  the two photons background by 
cutting on the two previous variables.

Another cut common to all the analysis is based on the polar angle of missing momentum 
and the effective centre of mass energy after initial radiation.
In the   $e^+e^-\ra q \overline{q}$ process the effective centre of mass energy can be 
considerably reduced  by the photon emission, and since the radiation is more likely 
along the beam pipe the missing momentum points at low polar angle.

\subsection{The likelihood and the Neural Network}
Several variables have different shape in the distribution in the phase space for signal and 
background. The informations given by several kinematic  distributions is combined using 
a likelihood technique or a Neural Network~\cite{garcia} and the output is a discriminant 
variable that allows to reach a high Higgs signal efficiency and high background rejection.
The two dimensional distribution of the discriminant variable and the 
mass  of the candidates   is used to give the signal over background ratio for each event. 
A test statistic is then constructed:
\begin{equation}
ln(Q)= -s_{TOT}+\sum_{i=1}^N n_i ln(1+\frac{s_i}{b_i})
\end{equation}
where $N$ is the total number of selected events, $s_{TOT}$ is the total signal rate and 
$\frac{s_i}{b_i}$ is the signal over background ratio for the event $i$.
The observed value of $ln(Q)$ is compared with the expected value from signal and background
experiments and the confidence levels of signal CL$_s$ and of background CL$_b$ 
are derived~\cite{anna}.

\section{$H\nu\overline{\nu}$ final state}
This is the second most likely final state. The energy flow and the jet reconstruction are 
two fundamental items to correctly reconstruct the Higgs mass. To improve the resolution
it is possible to impose  the recoil mass to be the $Z^0$ mass.
The drawbacks of this method are:
\begin{itemize}
\item in the events $e^+e^-\ra WW\nu_e\nu_e\ra H\nu_e\nu_e$ the recoil mass is wrongly 
reconstructed;
\item in the events at rest the mass is artificially reconstructed at the kinematic
limit $\sqrt{s}-m(Z)$ where the signal is expected, decreasing the signal over background ratio.
\end{itemize} 

The most difficult task is the treatment of the $q\overline{q}\gamma\gamma$ background.
In case each of the two photons is emitted symmetrically by the electron and positron
and they are lost in the beam pipe, the event is signal like: two visible jets in the final state 
with high invariant mass and high missing energy. The only possibility to discriminate
this background from the signal is to consider the acoplanarity  of the 
two jets: in the case of symmetric double radiative events, the two jets are nearly 
coplanar,
while in the signal case the events may be acoplanar even at the kinematic  limit
due to the $Z^0$ width and to the WW fusion contribution  to the Higgs production.    

Among the LEP candidates with higher signal over background ratio (see table~\ref{candi}) there are
3 $H\nu\overline{\nu}$ events.
The first has the two jets mass equal to 114.4 GeV/c$^2$, and a missing mass
of 94  GeV/c$^2$. It has a high value of b-tag and the acollinearity is 3 degrees.

\begin{table}[t]
\caption{Selected candidates with (s/b)$_{115} > 0.3$.\label{candi}}
\vspace{0.4cm}
\begin{center}
\begin{tabular}{|c|c|c|c|}
\hline
(s/b)$_{115}$ & M$_{rec}$   & Channel & Exp.\\
\hline
4.7 & 114 & Hqq & ALEPH\\
2.3 & 112 & Hqq & ALEPH\\
2.0 & 114 & H$\nu\overline{\nu}$ &  L3\\
0.9 & 110 & Hqq & ALEPH\\
0.6 & 118 & H$e^+e^-$ & ALEPH\\
0.52 & 113 & Hqq & OPAL\\
0.5 & 111 & Hqq & OPAL\\
0.5 & 115 & H$\tau^+\tau^-$ & ALEPH\\
0.5 & 115 & Hqq & ALPEH \\
0.49 & 114 & H$\nu\overline{\nu}$ & L3\\
0.47 & 115 & Hqq & L3 \\
0.45 & 97 & Hqq & DELPHI\\
0.40 & 114 & Hqq & DELPHI\\
0.32 & 104 & H$\nu\overline{\nu}$ & OPAL\\
\hline
\end{tabular}
\end{center}
\end{table}

\section{ Final state with charged leptons}
In these channels the background contamination can be  reduced to a smaller 
contribution than  is 
possible  for the other final states for the same signal efficiency.
Using constrained fits, the mass resolution can reach values of 2-3 GeV/c$^2$ and the typical 
problems of jet pairing of the four jets final states, or of energy flow of the missing energy 
channel are avoided. 
A particular care must be dedicated to the radiated photons from the high energetic leptons
in final state. If these photons are wrongly associated to the jets, they can artificially 
increase the two jets invariant mass and simulate a signal. 

An $He^+e^-$ and an $H\tau^+\tau^-$ are among the first 8  LEP candidates with highest 
signal over background ratio (see table~\ref{candi}). 

\begin{figure}[hbtp]
\begin{center}
\mbox{\epsfig{file=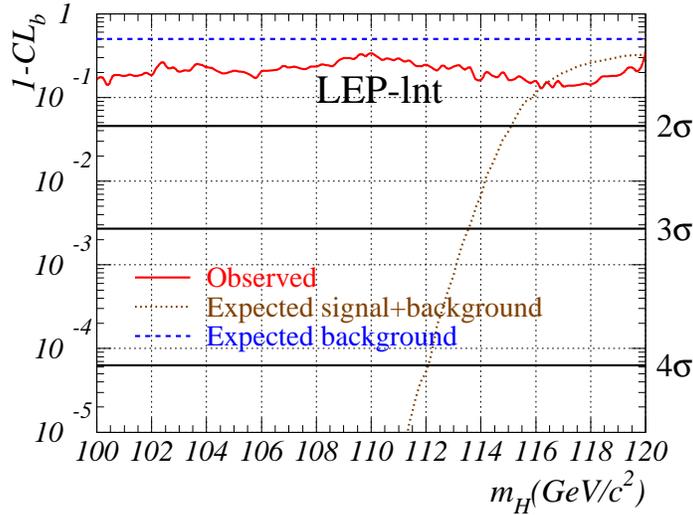,width=0.6\textwidth}}
\caption{$1-CL_b$ for missing energy and charged leptons final states.
}
\label{clb}
\end{center}
\end{figure}

\section{Conclusions}
The confidence level of the background for the  selected events in the missing energy and leptonic
channels can be extracted (see figure~\ref{clb}). There is an overall excess of events in data
with respect the expected events from Standard Model prediction, and this excess results
in a discrepancy of the order of one sigma from  the expected and 
observed confidence level of the background.

All LEP collaborations are reviewing their analysis taking into account the final calibration
of detectors, and more MonteCarlo statistic: the final word from LEP on Higgs will come in the 
forthcoming few months.

\end{document}